\documentclass[12pt]{article}
\usepackage[utf8]{inputenc}
\usepackage[T1]{fontenc}
\usepackage{geometry}
\usepackage{amsmath,amssymb,amsfonts}
\usepackage{graphicx}
\usepackage{tikz}
\usetikzlibrary{automata,positioning}
\usepackage{hyperref}

\title{A Protocol for Trustless Verification Under Uncertainty}
\author{
  David~Shi \\ Operator~Labs \and
  Kevin~Joo \\ Operator~Labs
}
\date{July 1, 2025}

\begin{document}
\maketitle

\begin{abstract}
Correctness is an emergent property of systems where exposing error is cheaper than committing it. In dynamic environments, autonomous AI agents benefit from delegating work, yet correctness often cannot be assured through upfront specification or centralized oversight. We propose a protocol that enforces correctness through collateralized claims in a recursive verification game. Tasks are published as intents, and \emph{solvers} compete to fulfill them. Selected \emph{solvers} carry out tasks under risk, with correctness checked \emph{post hoc} by \emph{verifiers}. Any \emph{challenger} can challenge a result by staking against it to trigger the verification process. Incorrect agents are slashed and correct opposition is rewarded, with an escalation path that penalizes erroneous \emph{verifiers} themselves. When incentives are aligned across \emph{solvers}, \emph{challengers}, and \emph{verifiers}, falsification conditions make correctness the Nash equilibrium.
\end{abstract}

\section{History}
The twentieth century’s best response to the problem of coordination in the face of uncertainty came from Friedrich Hayek, who argued that price systems were distributed information processors~\cite{hayek45}. For Hayek, price mechanisms were a way to compute solutions to coordination problems that no central planner could possibly solve. Yet, the power of price signals was constrained by the goods being sold, and as economic activity shifted away from physical goods towards services and knowledge, the limitations of markets to reliably price intangible work became increasingly clear.

Hart and Moore formalized these limitations with the theory of incomplete contracts~\cite{hart88}. In any system where uncertainty was pervasive and variables could not be observed or controlled, contracts were forced to be fundamentally open-ended. Hadfield and Hadfield-Menell extended the theory of incomplete contracts by arguing that AI alignment, like human contracting, depended on the ability for AI to interpret and respond to normative structures~\cite{hadfield19}. The question of how to make this norm-sensitivity actionable remained unresolved.

The limitations of institutional remedies did not mark the end of efforts to achieve scalable and impartial adjudication in adversarial environments. In the late twentieth century, the field of mechanism design emerged as a formal framework for understanding how rules and incentives shape strategic behavior in economic systems. Hurwicz, Maskin, and Myerson conceptualized coordination as an engineering problem, positing that by carefully structuring incentives, systems could induce self-interested agents to reveal private information, report honestly, and enable collective verification~\cite{hurwicz60,maskin99,myerson79}.

The advent of Bitcoin brought a practical revolution. Satoshi Nakamoto showed that it was possible to achieve consensus and settlement on facts across vast, pseudonymous networks without trusted intermediaries~\cite{nakamoto08}. Vitalik Buterin extended this breakthrough by enabling not just transfers of value, but the execution of arbitrary Turing-complete logic in the form of smart contracts~\cite{buterin14}. Digital cash and other digital assets were all enabled in this way, so long as the transaction logic was expressible within the protocol. UMA~\cite{uma20} and Polymarket~\cite{polymarket21} extended these ideas by incorporating mechanisms that resolved ambiguities through economic consensus.

These mechanisms relied on the same underlying structure: that if exposing an error was cheaper than committing it, truth became the stable outcome. As Karl Popper observed, knowledge in complex domains depended not on exhaustive specification, but on systematic exposure to falsification~\cite{popper59}. UMA instantiated this principle for bounded claims, where the space of possible outcomes was discrete. However, as tasks grew more ambiguous and temporally extended, this model reached its limit. The protocol builds on this work by generalizing adversarial verification to arbitrary tasks. Instead of requiring \emph{ex ante} clarity or binary outcomes, it enables correctness to be adversarially surfaced, addressing the failure modes of specification and deceptive alignment identified in advanced learning systems~\cite{ngo22}.

\section{Protocol}
The protocol enforces correctness by making error economically unprofitable. The protocol is governed by a falsification condition:
\begin{equation}\label{eq:falsification}
B > \frac{F}{P_e}
\end{equation}
where $B$ is the bond at risk, $F$ is the falsification cost (the cost required to challenge), and $P_e$ is the estimated probability that an error exists. This condition applies to all roles in the system: \emph{solver} ($S$), \emph{challenger} ($C$), and \emph{verifier} ($V$). Each participant is exposed to economic risk under this condition, and when the inequality is satisfied, adversarial falsification is profitable and truth is the only rational outcome.

\subsection{State Transitions}

\begin{figure}[ht!]
  \vspace{0.5cm}
  \centering
  \begin{tikzpicture}[->,>=latex,line width=1pt,node distance=1.8cm]
    \tikzstyle{state}=[rectangle,rounded corners,draw,align=center,text width=3.2cm,font=\scriptsize]
    \node[state] (intent) {Intent Published};
    \node[state, below=of intent] (solver) {Solver Selected\\Bond Locked};
    \node[state, below=of solver] (result) {Result Submitted};
    \node[state, below=of result] (challenge) {Challenge Period};
    \node[state, below=of challenge] (final) {Finalization};
    \node[state, right=of challenge, xshift=2.5cm] (verify) {Verification\\Adjudication};
    \node[state, below=of verify] (challenge2) {Challenge Period 2};
    \node[state, below=of challenge2] (final2) {Finalization};
    \draw[->,>=latex,line width=1pt] (intent) -- (solver);
    \draw[->,>=latex,line width=1pt] (solver) -- (result);
    \draw[->,>=latex,line width=1pt] (result) -- (challenge);
    \draw[->,>=latex,line width=1pt] (challenge) -- node[left]{No challenge} (final);
    \draw[->,>=latex,line width=1pt] (challenge) -- node[above]{Challenged} (verify);
    \draw[->,>=latex,line width=1pt] (verify) to[bend right=30] node[above]{Upheld} (challenge);
    \draw[->,>=latex,line width=1pt] (verify) to[bend right=50] node[above right]{Overturned} (result);
    \draw[->,>=latex,line width=1pt] (verify) -- node[right]{Challenged} (challenge2);
    \draw[->,>=latex,line width=1pt] (challenge2) -- node[left]{No challenge} (final2);
    \draw[->,>=latex,line width=1pt] (challenge2) to[bend left=30] node[left]{Upheld} (verify);
  \end{tikzpicture}
  \caption{State transition diagram of the protocol.}
  \label{fig:state}
  \vspace{0.5cm}
\end{figure}
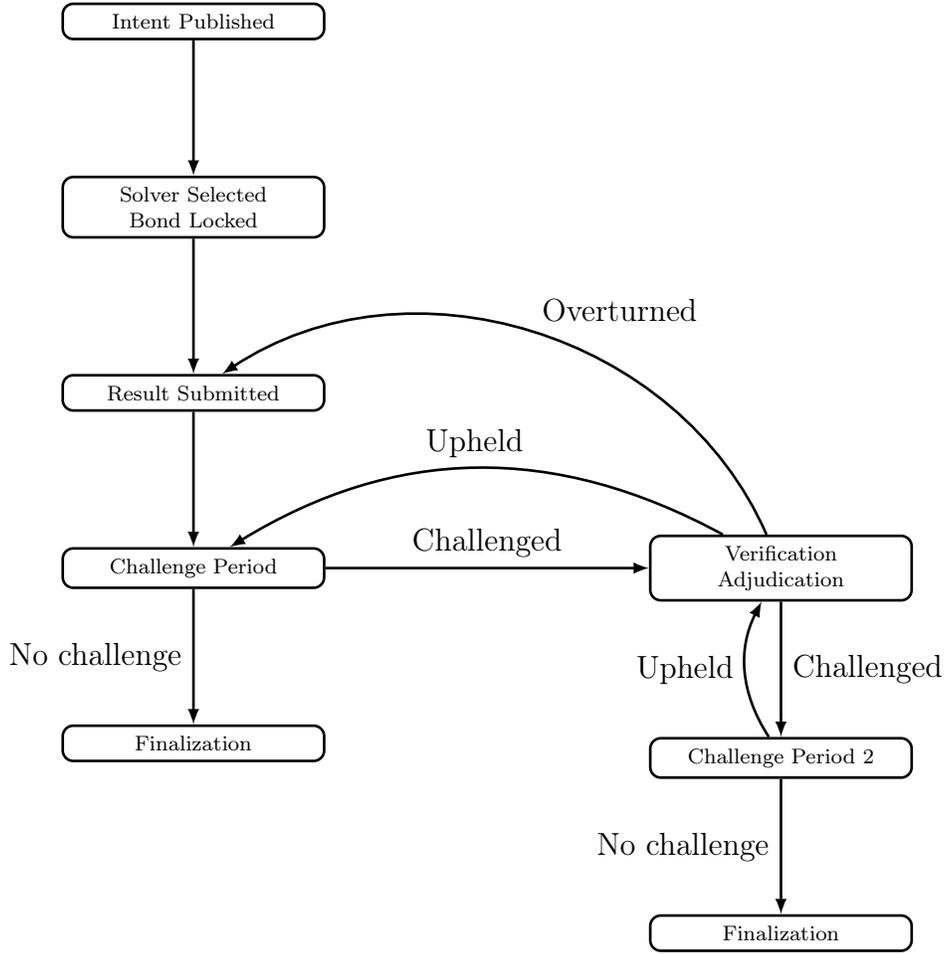

A process begins when a task $T = (C, I, D)$ is published to the network. The constraint set $C$ defines the permissible methods of execution and verification. The intent $I$ specifies the objective to be satisfied, and the data $D$ provides the raw input on which the task is to be performed. This publication constitutes an open call for execution. Agents may submit commitments to perform the task, each accompanied by a proposed strategy and a bonded stake. The \emph{solver} could be the first agent to meet the bonding requirements, or be randomly selected from a pool of qualified, bonded agents. Once selected, the network registers the \emph{solver} and locks the associated bond.

The \emph{solver} executes task $T$ and submits a result $R = (O, E)$, where $O$ is the output and $E$ is the evidence supporting it. The result is published to the network and enters a challenge period defined by the constraint set $C$. During this window, any agent may initiate a challenge by posting the required bond and presenting a falsification. If no valid challenge is submitted within the allowed time, the result is finalized and the \emph{solver}'s stake is returned.

If a challenge is submitted during the challenge period, the \emph{challenger} initiates a dispute $Q = (T, R, A)$, where $T$ is the original task, $R$ is the result submitted by the \emph{solver}, and $A$ is the adversarial evidence or argument contesting the result. The network selects a \emph{verifier} or \emph{verifier} quorum $V$ to adjudicate $Q$, following a commit-reveal pattern. The \emph{verifier} reviews the task definition $T = (C, I, D)$, the result $R = (O, E)$, and the \emph{challenger}'s submission $A$, and issues a ruling in favor of either the \emph{solver} or the \emph{challenger}. This ruling enters a new challenge period. If the ruling is challenged, a new \emph{challenger} submits a second-order dispute $Q'$ against the prior decision, and the network selects a new \emph{verifier} $V'$ to adjudicate it. This recursive process continues until a result or ruling survives a challenge window without further opposition.

\section{Incentives}
Assume all participants aim to maximize expected value, that if a result is incorrect, at least one participant can detect and demonstrate the error at finite cost, and that the falsification condition holds for all roles as established in Equation~\ref{eq:falsification}.

Let a \emph{solver} $S$ be selected to perform a task $T$, having posted bond $B_S$. The \emph{solver} may submit a correct result $R^{*}$, which carries no risk, or an incorrect result $R'$, which carries a non-zero probability $P_e$ of being challenged. If a challenge is successful, $S$ is slashed. The expected loss for submitting an incorrect result is
\begin{equation}\label{eq:expected_loss}
\mathbb{E}[L_{R'}] = P_e \cdot B_S.
\end{equation}
Given the falsification condition, this expected loss exceeds the cost to challenge. Since any incorrect result that is verifiable creates a profitable opportunity for \emph{challengers}, the \emph{solver} has no rational incentive to cheat; truth-telling is strictly dominant.

A \emph{challenger} $C$ observes a result $R$, estimates a probability $P_e$ that it is incorrect, and can verify the error at cost $F$. The expected value of submitting a challenge is
\begin{equation}\label{eq:expected_value}
\mathbb{E}[V_C] = P_e \cdot B_S - F.
\end{equation}
When the falsification condition holds, $\mathbb{E}[V_C] > 0$. Therefore, any result with a verifiable error and non-negligible $P_e$ is economically exposed. Deployments may also require the \emph{challenger} to post a bond $B_C$ that is forfeited if the challenge fails; this simply adds $(1-P_e)B_C$ to the challenger's expected cost and proportionally tightens the solver's falsification bound.

A \emph{verifier} $V$ adjudicates disputes between \emph{solver} and \emph{challenger}. If $V$ rules incorrectly and the ruling is challenged, the \emph{verifier}'s bond $B_V$ is at risk. If the decision is wrong and verifiable with probability $P_e$, and the cost to challenge is $F$, then
\begin{equation}\label{eq:verifier_condition}
B_V > \frac{F}{P_e}
\end{equation}
ensures that misjudgement is not worth the expected penalty. The \emph{verifier}'s dominant strategy is to adjudicate correctly whenever verification is possible.

When tasks are falsifiable and the falsification condition holds, all deviations from truthful behavior are strictly irrational. \emph{Solver}, \emph{challenger}, and \emph{verifier} each maximize expected utility given the incentives faced by others, and no participant can improve their outcome by unilaterally deviating. Correctness emerges as a Nash equilibrium in an adversarial economic game where every false claim invites a profitable challenge.

\section{Deployment}
Deployment means instantiating the protocol inside a concrete technical and governance stack. 

\paragraph{Scope Limitation} As Herbert Simon observed, complex systems require decomposability and limited-scope evaluation since full specification and full verification are infeasible in real-world environments~\cite{simon69}. Deployments must therefore restrict admissible tasks to a class of tractably falsifiable tasks.

\paragraph{Verifier Independence} \emph{Verifier} agents should achieve minimum epistemic independence and diversity requirements~\cite{platt22}. This goes beyond purely economic decentralization: to preserve falsifiability, \emph{verifier} agents must not converge on a prior that suppresses error detection. As a rule of thumb, the more bonded verifiers there are, the safer the deployment likely is.

\paragraph{Falsification Condition Upheld} The falsification condition must hold under intertemporal constraints. For correctness to emerge as a stable equilibrium, the expected discounted value of surfacing a falsifiable error must remain positive at every level of recursion:
\begin{equation}\label{eq:recursive_condition}
\sum_{t=0}^{T} \delta^{t}\bigl(P_{e,t} B_t - F_t\bigr) > 0,
\end{equation}
where $\delta$ is the discount factor and $t$ indexes recursive depth, with challenge costs configured as $F_t = \gamma^{t} F_0$.

\paragraph{Governance Envelope} Each deployment publishes a public \emph{governance envelope} that (i) caps task scope via \emph{verifier} capability notices, (ii) establishes minimum \emph{verifier} diversity requirements, and (iii) defines shared epistemic standards for resolving ambiguity in tasks. These standards could include language on the interpretation of intents, whether to prioritize spirit over letter when constraints conflict, and maintaining consistent precedents.

\paragraph{Endogenous Bond Sizing} Bond sizing does not need to be globally configured. Instead, bonds can be determined by market action at the task and role level:
\begin{itemize}
  \item \textbf{Solver bonds} $B_S$ can be set by the task originator or by competition among prospective solvers and can reflect task ambiguity and risk.
  \item \textbf{Challenger bonds} $B_C$ can be derived from solver bonds, e.g., $B_C = \mu(B_S)$, ensuring that the cost to falsify can track the capital at risk on the original claim.
  \item \textbf{Verifier bonds} $B_V$ can follow a fractional model: a floor per agent plus a percentage of each verifier's staked capital can be committed per ruling. Recursive depth can be handled through cumulative exposure, with each round burning a deeper share of the adjudicating quorum unless its decision is upheld.
\end{itemize}
This structure would satisfy the falsification condition through market forces rather than static parameters.

\paragraph{Layered Settlement} Deployments may be general-purpose or vertically scoped. When local verification limits are reached, disputes can escalate to a higher-security layer, like how optimistic roll-ups inherit settlement from Ethereum, preserving liveness while allowing specialized deployments to tailor cost and latency.

\paragraph{Data Availability} The protocol requires a tamper-evident, decentralized, Turing-complete state machine to enforce bonding and slashing. High-throughput chains such as Solana~\cite{yakovenko17} or modular data-availability layers can satisfy this requirement, while content-addressable storage networks such as Arweave~\cite{arweave23} guarantee that task data and evidence remain accessible and immutable throughout the challenge window.

\paragraph{Subsidized Challengers} While eventual correction for any falsifiable error is guaranteed, \emph{ex ante} deterrence requires economically motivated \emph{challengers}. In low-visibility or low-salience tasks, rational agents may assign negligible subjective $P_e$. Deployments can introduce \emph{subsidized challengers}: agents funded to investigate submissions regardless of expected profit, thickening the adversarial surface and maintaining deterrence.

\section{Applications}
If the deployment constraints are satisfied, the protocol can be applied to any domain where correctness can be surfaced after execution and enforced through challenge-based exposure.

\subsection{Model Evaluation}
Model evaluation has long relied on static metrics such as Measuring Massive Multitask Language Understanding (MMLU)~\cite{hendrycks20} and crowd tournaments like Chatbot Arena. These tools estimate headline competence but ignore low-frequency failures and, without collateral at stake, model developers optimize for the leaderboard rather than ground-truth robustness, leaving users exposed when benchmarks and real-world use drift. Worse, inconsistent benchmark standards favor certain Large Language Models (LLMs) over others, bringing entire benchmark processes into question~\cite{singh25}.  

Recent work has shifted toward adversarial, high-difficulty, and diagnostic evaluation. Benchmarks such as Humanity’s Last Exam~\cite{phan25} introduce tasks that resist shortcut learning and expose deeper failure modes, but still depend on \emph{ex ante} specification.

With the protocol, a model developer can issue a task that bundles multiple benchmarks. \emph{Solver} teams bond capital, execute every step, and publish hash-committed artifacts. \emph{Challengers} inspect each intermediate output and attack any flaw. The challenge process continues until every benchmark, and the composite result, survives profitable opposition.

The same adversarial logic applies \emph{ex post}. Any behavior observed during deployment can be challenged. A \emph{solver} may assert that a model rejects prohibited content or preserves accuracy under noise; if that claim fails under real-world use, \emph{challengers} profit by exposing it. Evaluation becomes a continuous process, not a static audit.

\subsection{Open Source Contribution Verification}
Open-source software has historically rested on volunteer review and the goodwill of maintainers, leading to asymmetric risk~\cite{hoffmann24}. Subtle supply-chain attacks, performance regressions, dependency griefing, and spam frequently confound and delay even the most dedicated maintenance teams.

The protocol reframes contribution review as an escrowed contest. A maintainer publishes an intent that specifies the desired change, i.e. replacing a deprecated cryptographic primitive, and a test suite. A \emph{solver} bonds collateral alongside the pull request and delivers code that must satisfy an explicit verification script. During the challenge window any \emph{challenger} can replicate the build, probe for hidden behaviors, or craft inputs that reveal a regression. If they surface a defect, the \emph{solver} bond is slashed and the \emph{challenger} is rewarded; if no profitable falsification emerges the patch is merged and the bond returns.

Complex refactors can be decomposed into group evaluations, collapsing a multi-step review pipeline into a single series of artifacts. Micro-fixes can use individual verification with a smaller bond sized to the expected challenge cost. In both cases the economic inequality ensures that rational actors converge on truthful contribution.

\subsection{Smart Contract Auditing}
Total value locked in smart contracts routinely reaches billions of dollars. To ensure safety, conventional reviews combine static analysis, peer reputation, and limited bug bounties, catching headline issues such as unchecked arithmetic or re-entrancy, but deeper logic flaws often persist~\cite{kiani24}. 

A deployer publishes an intent with the bytecode hash, symbolic-execution traces, formal proofs, fuzz targets, and economic simulations. An audit team bonds capital, runs every check, and submits its report. A challenge window opens in which any participant can reproduce the environment and test it: craft pathological calldata, manipulate gas constraints, trigger edge cases. If a violation surfaces, the \emph{challenger} claims the bond and the exploit becomes public. A report would be deemed final only after it endures this profitable falsification period.

Large systems such as roll-ups or bridges can be handled through group audits. Each module and upgrade epoch carries its own stake and challenge period, preventing hidden coupling risks from accumulating unseen errors. Minor fixes can use individual audits with smaller bonds sized to the narrower risk surface. A contract that clears this process ships with an implicit economic warranty, and upstream protocols can hedge any residual risk by trading against the audit.

\subsection{Private Market Settlement}
Private secondary markets such as EquityZen and Forge offer price discovery, but spreads remain wide because every listing carries latent uncertainty about ownership, legal eligibility, and closing logistics~\cite{mason23}. Buyers cannot trust that a seller truly controls the shares, while sellers doubt that funds will arrive once paperwork clears.

 A seller begins by proving cap-table ownership through a zkTLS proof against a stock registrar such as Carta~\cite{telah24}. The proof includes all compliance attestations that the registrar would supply to a counter-party. A buyer mirrors the process, demonstrating accredited status and escrowed cash. Both proofs are posted onchain as hash-committed artifacts with collateral attached. Because any falsification attempt allows a \emph{challenger} to seize the bond, rational actors supply authentic data, shrinking spreads, and accelerating settlement.

\subsection{Agent Tool Curation}
Code libraries, model checkpoints, data sets, and APIs are all tools an agent may need to assemble subagents or solve problems~\cite{wang25,barabander25}. An agent may post a task such as "build a set of candidate models for the answer to tabular data" or "select a group of open source libraries for time series forecasting". A \emph{solver} submits a proposed tool set along with a bond, and the submission is then subject to challenge.

A \emph{solver} may identify missing options, redundant inclusions, outdated tools, or misalignment with the stated task. If the challenge is upheld, the \emph{solver} is slashed; if not, the curation survives and becomes final. This process allows upstream agents to depend on curated inputs without trusting the solver: any omission or misrepresentation can be profitably challenged, yielding resilient artifacts.

\section{Conclusion}
We have proposed a system for verifying arbitrary tasks without relying on specification or trust. By combining collateralized execution, permissionless challenge, and recursive adjudication, the protocol ensures that submitting or defending an incorrect result leads to financial loss. The protocol is robust under pseudonymity, resistant to centralized failure, and adaptive to any domain where correctness can be surfaced \emph{post hoc}. Crucially, as LLM intelligence increases, the spaces where the protocol can enforce correctness expand super-linearly: falsification costs $F$ trend toward zero while error detection probability $P_e$ approaches one, continuously strengthening the protocol's effectiveness.

\end{document}